\documentclass[journal]{IEEEtran}

\usepackage{amsmath}
\usepackage{amssymb}
\usepackage{xfrac}
\usepackage{array}
\usepackage{booktabs}
\usepackage{url}
\usepackage{cite}
\usepackage{cleveref}
\usepackage[caption=false]{subfig}
\usepackage{graphicx}
\usepackage{multirow}
\usepackage{makecell}
\usepackage{array}
\usepackage[table]{xcolor}
\usepackage{tikz}
\usepackage{siunitx}
\usetikzlibrary{shapes.geometric, arrows}
\usepackage{comment}
\usepackage[utf8]{inputenc}
\usepackage[T1]{fontenc}

\newcommand\submittedtext{%
  \footnotesize This work has been submitted to the IEEE for possible publication. Copyright may be transferred without notice, after which this version may no longer be accessible.}

\newcommand\submittednotice{%
\begin{tikzpicture}[remember picture,overlay]
\node[anchor=south,yshift=10pt] at (current page.south) {\fbox{\parbox{\dimexpr0.65\textwidth-\fboxsep-\fboxrule\relax}{\submittedtext}}};
\end{tikzpicture}%
}

\begin{document}
\title{Towards Accurate MHz Core Loss Measurements}
\author{Nick~J.~Kirkby,~\IEEEmembership{Student~Member,~IEEE,}
        and~Mike~K.~Ranjram,~\IEEEmembership{Member,~IEEE}%
\thanks{The authors are with the School of Electrical, Computer, and Energy Engineering, Arizona State University, Tempe, AZ 85281 USA (e-mail: nkirkby@asu.edu; mranjram@asu.edu).}
}

\maketitle

\submittednotice

\begin{abstract}
  Miniaturized magnetic components play an important role in achieving high efficiency and high power density in power converters.
  Accurate core loss data are essential when optimizing magnetic components, 
  but making accurate core loss measurements remains challenging, especially at MHz frequencies. 
  As a result, many research groups have developed their own testers in pursuit of improved accuracy, 
  and the amount of published core loss data has increased.
  However, large variations in these data shed light on fundamental barriers to accurate MHz core loss measurement.  
  This paper identifies these barriers and presents methods for overcoming them.

\end{abstract}

\IEEEpeerreviewmaketitle

\section*{Nomenclature}

\begin{table}[h!]
\centering
\begin{tabular}{cll}
  \toprule
Symbol      & Description & Defined by \\ 
\midrule
$P_V$       & Volumetric core loss                  & \multirow{3}{*}{IEC 62044-3:2023} \\
$\hat{B}_e$ & Peak effective magnetic flux density  & \\
$\mu_{ea}$  & Effective amplitude permeability      &  \\ \hline
$l_e$       & Effective length                      & \multirow{3}{*}{IEC 60205:2016} \\ 
$A_e$       & Effective cross-sectional area        & \\ 
$V_e$       & Effective volume                      & \\ \hline
$u(x)$      & Standard uncertainty of $x$           & \multirow{2}{*}{JCGM 100:2008} \\
$k$         & Coverage factor                       &  \\
\bottomrule
\end{tabular}
\end{table}


\section{Introduction}
\begin{figure}[t]
  \centering
  \begin{tikzpicture}
    \node (base) [inner sep=0pt] {
      \input{data_comparison/plots/core_loss/pgf/PowerLoss_DMR51W_2000kHz.pgf}
    };
    \node (overlay) at (base.center) [xshift=1.21cm, yshift=0.02cm, inner sep=0pt] {
      \includegraphics[width=0.26\textwidth]{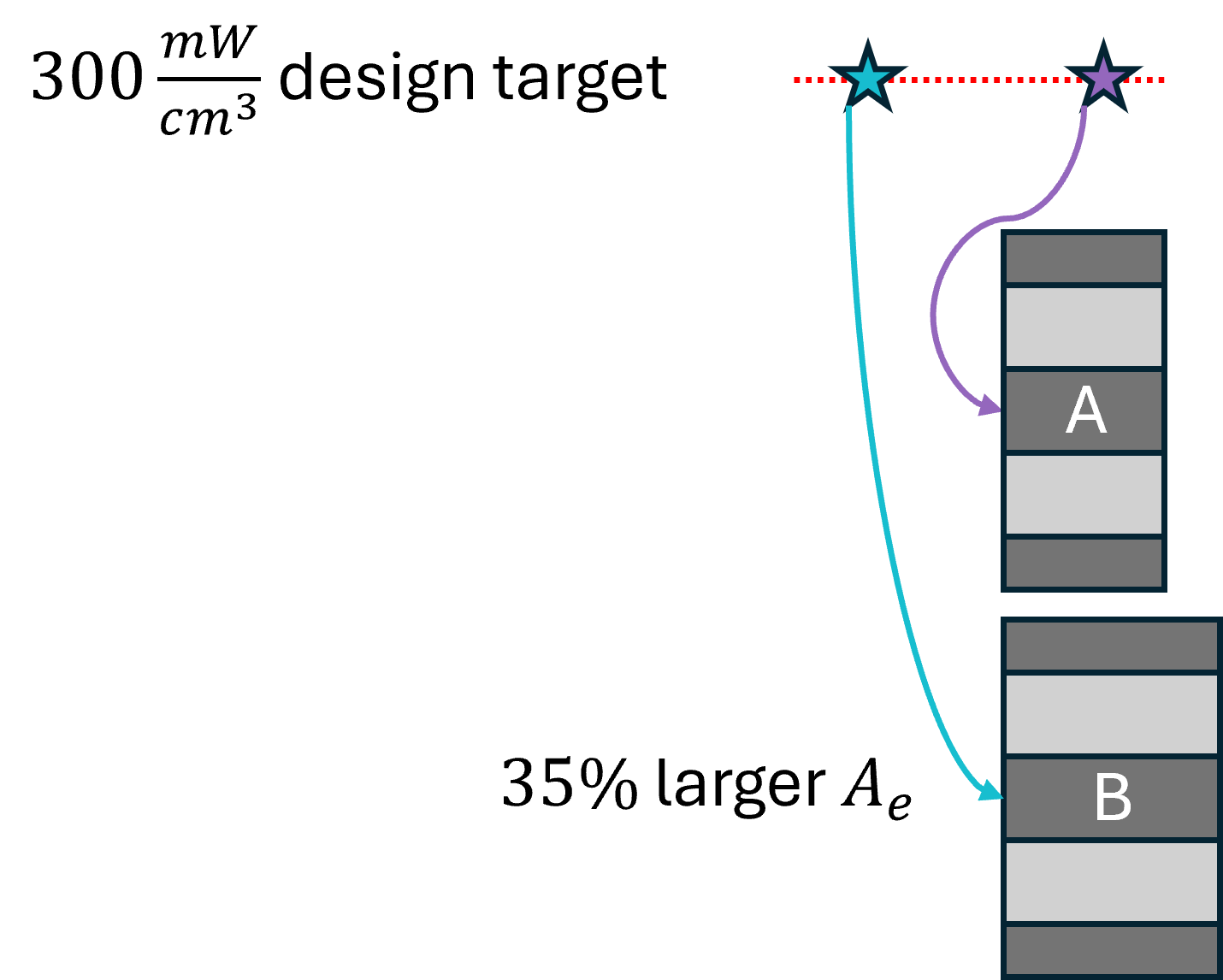}
    };
  \end{tikzpicture}
  \caption{Differing core loss measurements of DMR51W material at 2 MHz by This Work and two other independent laboratories \cite{nabih_core_2026,dmegc_dmr51w-revb_material_characteristics_2018}. 
  The effect of using \cite{dmegc_dmr51w-revb_material_characteristics_2018} versus \cite{nabih_core_2026} in design of an E core for a 300 $mW/cm^3$ design target are shown as "A" and "B" respectively.  }
  \label{fig:DMR51W_comparison}
\end{figure}

\IEEEPARstart{O}{ne} decade has passed since the publication of \cite{hanson_measurements_2016} in TPEL, 
which showed that magnetic component power density improvements of 3-10x are possible by operating loss-limited ferrites at 1-10 MHz, 
compared to more typical saturation-limited operating frequencies of 100 kHz or lower. 
The promise of miniaturized magnetic components has motivated much research aimed at practical realization of high frequency switching converters, 
including:
high frequency ac winding loss mitigation
\cite{fei_high-efficiency_2017,yang_low_loss_2019,yang_design_2021,joisher_high-performance_2025,ranjram_380-12_2022,willows_multi_level_2024},
improved modeling of high frequency winding terminations and resonances 
\cite{prakash_termination_2023,ranjram_design_2019},
study of electromagnetic wave effects in ferrite 
\cite{nabih_high_2021,kacki_analysis_2023},
and modeling of switching device losses and parasitics
\cite{zhuang_small-_2021,zulauf_impact_2020,zulauf_active_2019,zulauf_coss_2018}.
However, harnessing these valuable findings for miniaturization of magnetics remains challenging. 
In particular, optimized designs which aim to minimize volume or footprint such as \cite{ranjram_380-12_2022,nabih_high_2021,wouters_field-balanced_2026} 
are sensitive to core losses, 
and variations in the data used to fit core loss models have significant impact on the resulting design optima. 
For example, \Cref{fig:DMR51W_comparison} shows measurements of DMR51W cores made at 2 MHz by three independent laboratories under nominally identical conditions\footnote{
All three sets of measurements are made at 2 MHz, sinusoidal waveshape, 25C temperature, zero dc bias, and nominal T12.5/7.5/7 core size.
}, but the three measurements differ, and this difference changes the optimal design.
Using data from \cite{nabih_core_2026} instead of \cite{dmegc_dmr51w-revb_material_characteristics_2018} 
results in a design having 35\% larger core area (and volume) to achieve the same loss density target of $300 \frac{mW}{cm^3}$.
This variation in measurements makes it challenging to confidently assess a-priori whether a designed component will meet efficiency and volume requirements.
Determining the true optimal design entails constructing multiple hardware variants, costing time and money which may not be practical or available.

The lack of error bounds for core loss data is a significant obstacle to realizing the promise of miniaturized magnetic components.
This stands in contrast to other aspects of power converter design such as selecting power semiconductor devices, 
for which typical and maximum values of key metrics like $R_{ds,on}$ are routinely quoted in datasheets.
However, power magnetic material datasheets only provide nominal values for core loss with no indication of maximum value, 
posing a major challenge for designers of magnetic components.
Although research on MHz core loss measurement spans multiple decades 
\cite{bady_measurement_1958,goldberg_high_1988,gradzki_core_1992,skutt_high-frequency_1996,prabhakaran_impedance-analyzer_2002,han_evaluation_2012,syed_high_2013,mu_high_2013,papamanolis_transient_2021,princeton_power_electronics_magnet_data_quality_webinar_2023,anderson_automation_2023,brown_parallel_2024,anderson_class-d_2024},
measurement uncertainty remains difficult to quantify.
Commercially available MHz core loss test equipment does not state measurement uncertainty \cite{iwatsu_2017,mindcet_madmix_2026}, 
and rigorous error analyses are only relatively recent developments in the research community \cite{papamanolis_transient_2021,princeton_power_electronics_magnet_data_quality_webinar_2023}.
These recent analyses have identified fundamental issues with present measurement practices,
and explain the core loss accuracy problems that broadly affect power magnetic designers today.

The first fundamental issue is error modeling.
The Princeton MagNet team in \cite{princeton_power_electronics_magnet_data_quality_webinar_2023} emphasized that measurement error 
must be reported if data gathered on different testers is to be meaningfully compared.
In addition, quantifying reproducibility in data comparisions requires that all parties report measurement error \cite{ISO13528:2022}.
However, measurement error for core loss testers remains largely unreported.
Even though IEC 62044-3, the standard for core loss measurement, 
mandates that measurement error must be quantified and included in test reports,
the authors have been unable to find any manufacturer datasheet or commercial core loss tester\cite{iwatsu_2017,mindcet_madmix_2026} which reports measurement error.
To address this issue, this paper presents a simple and versatile error modeling framework suitable for all types of core loss testers in \Cref{sec:guide}.


The second fundamental issue is the identification and modeling of non-ideal \textit{effects} inherent to the practical realization of a measurement principle, 
such as circuit parasitics.
According to \cite{jcgm_gum_6_2020}, effect modeling is one of the most demanding tasks in measurement system development.
\cite{papamanolis_transient_2021} found that electrical parasitic effects in electrical testers are a barrier to accurate measurement at MHz frequencies, 
which led them to develop a novel transient calorimetric measurement principle.
Indeed, the prevailing approach in electrical core loss testers in \cite{han_evaluation_2012,mu_high_2013} is to 
assume that circuit parasitics have negligible influence on the measured quantities, but this may not always be the case.
Explicitly compensating for circuit parasitics is rare due to model complexity and challenges with validation as in \cite{syed_high_2013}.
We share methods for explicit effect modeling in \Cref{sec:effect_models} which make challenging effects more tenable.
In addition, we propose a novel calibration scheme for resonant core loss testers in \Cref{sec:component_stage_calibration}
which improves confidence in the correctness of the measured quantities.



\begin{figure}
  \centering
  \input{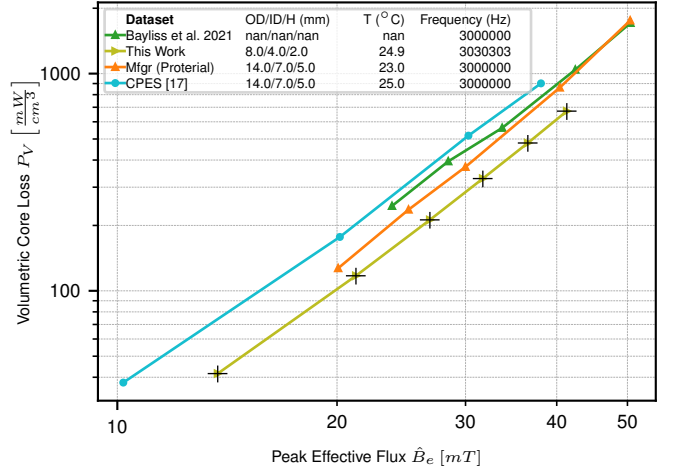}
  \caption{Proterial ML91S core loss measurements at 3 MHz by \cite{bayliss_design_2021,proterial_mn-zn_datasheet_2022,nabih_core_2026} and This Work.
  All surveyed data are made available as .csv in \cite{Kirkby_MHz_Core_Loss_2026}.
  }
  \label{fig:ML91S_3MHz_comparison}
\end{figure}

A third problem is that ``ground truth'' \cite{princeton_power_electronics_magnet_data_quality_webinar_2023} does not exist for $P_V$ or $\hat{B}_e$,
meaning core loss testers cannot be calibrated against a known value.
In contrast, other measurement equipment such as oscilloscopes or voltmeters are calibrated using readily available transfer standards 
which are traceable via calibration chain back to fundamental physical constants.
Traceable transfer standards make it possible to quickly check the accuracy of measurement equipment, 
but such standards do not yet exist for core loss measurement systems.
As a result, when faced with conflicting core loss data such as those in \Cref{fig:ML91S_3MHz_comparison}, 
it is difficult to know exactly why the data disagree.

Lastly, the question of measurement reproducibility remains unanswered.
Establishing reproducibility is a key aim of recent interlaboratory studies by ScN Magnetics using the Magnetic Core Evaluation Kit \cite{martinez_magnetic_2022,friebe_core_2025},
a controlled set of cores which is passed between laboratories for comparison.
Accuracy and stability of measurements are both challenges identified in \cite{rylko_magnetic_2026}, 
and reproducibility cannot be determined unless participants report measurement error, which most do not.
In addition, current ScN studies are focused on the 10-200 kHz frequency range, 
and MHz operation is anticipated to bring additional challenges.

\begin{figure}
  \centering
  \includegraphics[width=0.48\textwidth]{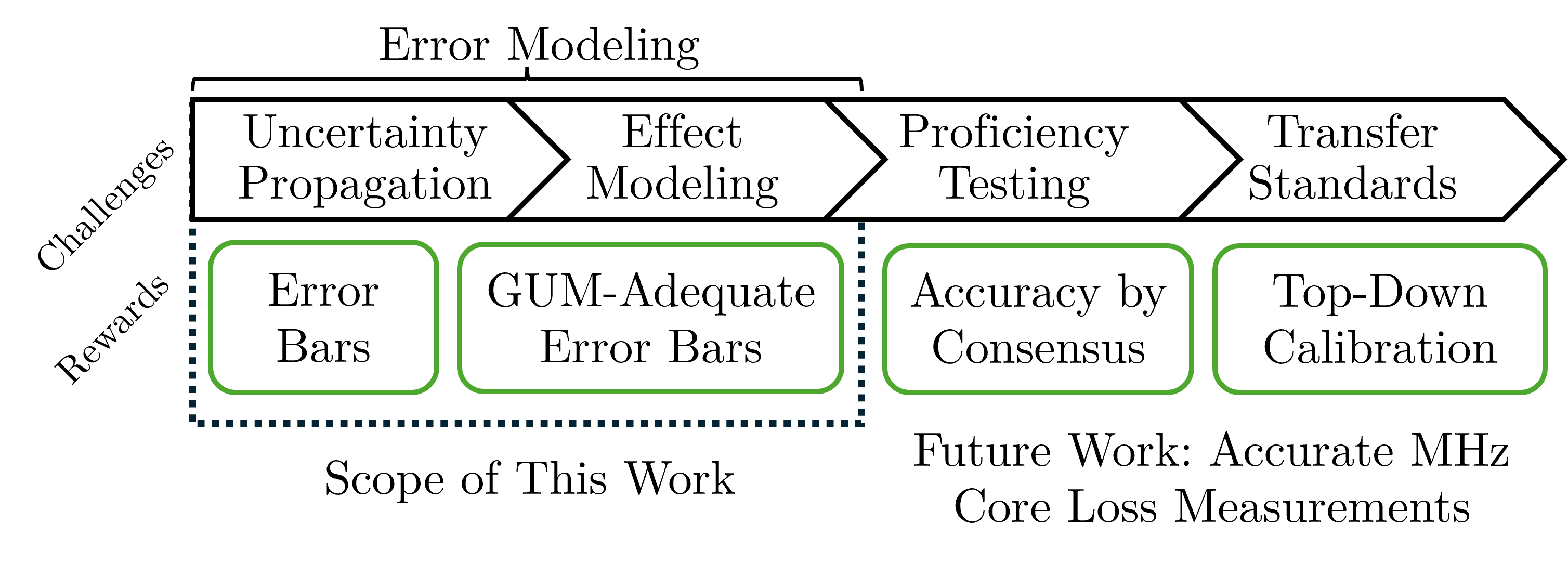}
  \caption{Challenges and rewards on the way to accurate MHz core loss measurements.}
  \label{fig:barriers_and_rewards}
\end{figure}

These four, intertwined barriers of error modeling, challenging effects, ground truth, 
and reproducibility block the road to accurate MHz core loss measurements.
This paper proposes a way forwards, summarized in \Cref{fig:barriers_and_rewards}, 
starting with error modeling.
As noted in \cite{princeton_power_electronics_magnet_data_quality_webinar_2023},
clear and consistent communication of measurement uncertainty is key to progress.
As such, a key contribution of this paper is the simple error modeling method for core loss testers presented in \Cref{sec:guide}.
\Cref{sec:component_stage_calibration} builds on this method, 
describing how effect modeling can be used to bring component-stage measurands into agreement with example calibration data. 
\Cref{sec:effect_models} addresses effects modeling for core-stage measurands. 
\Cref{sec:mm_aided_development} describes how the proposed methods in this paper can be used to identify and reduce uncertainty. 
\Cref{sec:future_work} proposes future work that can be conducted by the research community to build on this paper towards accurate MHz core loss measurements. 
Finally, \Cref{sec:conclusion} concludes the paper.


\section{Guide to Uncertainty Propagation for Core Loss Testers}
\label{sec:guide}

This section demonstrates a simple method of uncertainty propagation for core loss testers. 
Uncertainty propagation is the method used to compute measurement uncertainity (error) for the measured quantities.
The method presented here is based on the Guide to Uncertainty in Measurement (GUM) \cite{jcgm_gum_2008,jcgm_gum_102_2011,jcgm_gum_6_2020},
the foremost internationally standardized metrological guide.
The method is applicable to all core loss measurement principles, including volt-amperic, resonant, and calorimetric.
In addition, the method is easily retrofitted to existing testers -- it does not require any hardware changes.
For simplicity, this guide omits effect models, which are introduced in \Cref{sec:component_stage_calibration,sec:effect_models}.
Later, effect models are incorporated as additional terms,
but the procedure for computing measurement uncertainty remains unchanged.
Although the method is applicable to all measurement principles, 
the examples are based on the series-resonant electrical principle described in \cite{anderson_automation_2023,anderson_class-d_2024} 
and shown in \Cref{fig:QMeter1} for added clarity of working through a concrete example.


\subsection{Measurement Model Definition}

The measurement model is a compact notation for relating inputs and outputs of a measurement system, standardized in \cite{jcgm_gum_102_2011}, 
and shown as $\textbf{f}$ in \labelcref{eq:measurement_model_general},
\begin{equation}
\textbf{Y} = \textbf{f}(\textbf{X})
\label{eq:measurement_model_general}
\end{equation}
where $\textbf{Y}$ is the vector of measurands (outputs) and $\textbf{X}$ is the vector of inputs.
This notation has three advantages:
First, it supports multiple measurands, 
which is needed for all core loss testers since $P_V$ and $\hat{B}_e$ are both subject to uncertainity.
\footnote{
As yet, other publications on core loss error modeling \cite{papamanolis_transient_2021,princeton_power_electronics_magnet_data_quality_webinar_2023}
consider uncertainty in $\hat{B}_e$ to be negligble, 
but we find in \Cref{sec:mm_aided_development} that it is significant.
}
Second, the standard form makes it compatible with existing GUM-based measurement uncertainty analysis software such as \cite{delker_suncal_2024,NIST_Uncertainty_Machine_Possolo2013}.
Third, it is easy to adopt, 
because existing MHz core loss testers \cite{han_evaluation_2012,mu_high_2013,papamanolis_transient_2021,anderson_automation_2023,brown_parallel_2024} have already derived equations to compute $P_V$ and $\hat{B}_e$,
so rearrangement into standard form is the only requirement to complete this step.

\begin{figure}
  \centering
  \includegraphics[width=0.48\textwidth]{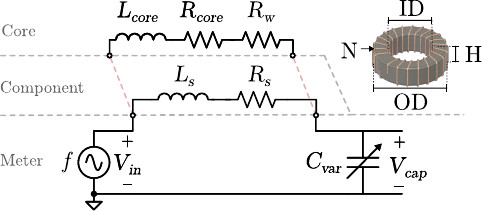}
  \caption{Ideal (parasitic-free) Q-Meter circuit showing core and component-stage models,
  used as an example measurement principle for conducting uncertainty propagation in this paper.}
  \label{fig:QMeter1}
\end{figure}

For example, a measurement model based on the series-resonant principle in \Cref{fig:QMeter1} and described in \cite{anderson_class-d_2024} 
is shown in \labelcref{eq:meas_model_compact},
\begin{equation}
\begin{aligned}
\textbf{Y} = [ R_s, L_s, P_V, \hat{B}_e, \mu_{ea} ]^\intercal = \textbf{f}(\textbf{X})
\end{aligned}
\label{eq:meas_model_compact} 
\end{equation}
where $\textbf{Y}$ contains two component-stage measurands, 
series resistance $R_s$ and inductance $L_s$, 
and three core-stage measurands, volumetric core loss $P_V$, peak effective flux $\hat{B}_e$, 
and effective amplitude permeability $\mu_{ea}$.
The measurement functions $\textbf{f}(\textbf{X})$ are listed in \Cref{sec:appendix_mm_35}, 
and the nine inputs $\textbf{X}$ are shown in \labelcref{eq:meas_model_X},
\begin{equation}
\begin{aligned}
\textbf{X} = [V_{cap}, V_{in}, C_{var}, f, R_{w}, N, OD, ID, H]^\intercal
\end{aligned}
\label{eq:meas_model_X} 
\end{equation}
\Cref{fig:QMeter1} shows the circuit-domain model used to derive $\textbf{f}(\textbf{X})$,
where frequency and voltage of the oscillator are $f$ and $V_{in}$ respectively, 
and the value of the variable capacitor is $C_{var}$.
The remaining inputs $X_{5...9}$ are the winding resistance $R_w$,
the number of turns $N$, 
and the toroid dimensions: outer diameter $OD$, inner diameter $ID$, and height $H$. 
Nominal values $x_i$ of each of the inputs are shown in \Cref{tab:input_values_and_uncertainties}.

\subsection{Input Uncertainty Budgeting}
\label{sec:guide_input_uncertainty}

In this step, the uncertainty of each input is estimated.
When first setting up an error model, large and coarse estimates are sufficient to gain intuition about system performance.\footnote{
Later, these estimates can be reduced and refined through input calibration or detailed characterization. 
Since the time and capital costs of these refinements may be substantial, 
it is best to use the results of a coarse uncertainty propagation to justify subsequent targeted refinements.
}
In our case, manufacturer-specified uncertainty bounds were available for many of the inputs, 
and these bounds are shown in \Cref{tab:input_uncertainties}.
\begin{table}
\caption{Uncertainty Bounds specified by Equipment Manufacturers}
\label{tab:input_uncertainties}
\centering
\begin{tabular}{| l | l | c |}
\hline
\textbf{Input} & \textbf{Bounds} & \textbf{Equipment} \\
\hline
$C_{var}$ & $\pm1\%$ & Meiden SCV-510H65W-705E \\
$OD,ID,H$ & $\pm 0.02\ mm$ & Mitutoyo CD-6 CSX \\
$R_{w,dc}$ & $\pm 0.05\%$ & GW-Instek LCR-819 \\
$V_{cap},V_{in}$ & $\pm 1\%$ & Tektronix MSO46B \\
$V_{cap}$ & $\pm 2.2\%$ & Tektronix TPP0850  \\
$V_{in}$ & $\pm 2.2\%$ & Tektronix TPP1000 \\
\hline
\end{tabular}
\end{table}
However, 
Some inputs may have multiple uncertainty components associated with them. 
For example, $V_{in}$ has $\pm 1\%$ uncertainty due to the oscilloscope and another $\pm 2.2\%$ uncertainty from the voltage probe attenuation ratio.
In addition, uncertainity is expressed in a variety of ways. 
The manufacturer-specified bounds in \Cref{tab:input_uncertainties} are stated as either relative (\%) or absolute values, 
however, uncertainty may also be communicated as a probability distribution.
To do propagation, these uncertainties must be combined and homogenized such that each input has just one uncertainty which represents all components which affect that input.
This procedure, sometimes called uncertainty budgeting, 
can be partially automated by software packages such as SUNCAL\cite{delker_suncal_2024} used here\footnote{
For completeness, a summary of the methods used to combine and homogenize uncertainity components is listed in \Cref{sec:appendix_converting_and_combining_uncertainties}.
}, and the resulting standard uncertainties are listed in the $u(x_i)$ column of \Cref{tab:input_values_and_uncertainties}.
\begin{table}
\centering
\caption{Input Values and Uncertainty Components at One Operating Point of the ML91S Sample }
\label{tab:input_values_and_uncertainties}
\begin{tabular}{c c c c l}
\hline
$X_i$ & $x_i$ & $u(x_i)$ & Unit  \\ \hline
$OD$ & 8.00 & 0.012 & $\mathrm{mm}$ \\
$ID$ & 4.00 & 0.012 & $\mathrm{mm}$ \\
$H$ & 2.00 & 0.012 & $\mathrm{mm}$ \\
$N$ & 14 & 0 & --- \\
$C_{var}$ & 108.20 & 0.62 & $\mathrm{pF}$ \\
$R_{w}$ & 0.91644 & 0.00026 & $\Omega$ \\
$V_{cap}$ & 37.69 & 0.53 & $\mathrm{V}$ \\
$V_{in}$ & 2.656 & 0.037 & $\mathrm{V}$ \\
$f$ & $2.00\times10^{6}$ & 0.0 & $\mathrm{Hz}$ \\ \hline
\end{tabular}
\end{table}

\subsection{Propagation with SUNCAL}


Uncertainty is propagated using the software package SUNCAL\cite{delker_suncal_2024}.
A configuration file \texttt{ideal.yaml} is provided in the supplemental materials, 
which contains all parameters needed to replicate the results shown in \Cref{tab:uncertainty_results}.
The table shows the nominal measurand estimate $y_j$, 
the combined uncertainty $u_c(y_j)$, 
and the expanded relative uncertainty $U_r$ for each of the measurands of the component and core stages.
\begin{table}[t]
\centering
\caption{Measurand Values and Combined Uncertainty for ML91S sample at \Cref{tab:input_values_and_uncertainties} Operating Point}
\label{tab:uncertainty_results}
\begin{tabular}{l l l l l c}
\hline
Stage & $Y_j$ & $y_j$ & $u_c(y_j)$ & Unit & $U_r(y_j)$ \\ \hline
\multirow{2}{*}{Component} & $R_{s}$ & 51.8 & 1.1 &  $\Omega$ & 4.25\% \\
& $L_{s}$ & 58.5 & 0.34 &  $\mu\mathrm{H}$ & 1.16\%  \\ \midrule
\multirow{3}{*}{Core} & $\hat{B}_{e}$ & 55.74 & 0.87 &  $\mathrm{mT}$ & 3.12\% \\
& $P_{V}$ & 999 & 22 &  $\mathrm{mW}\,/\,\mathrm{cm}^{3}$ & 4.40\% \\
& $\mu_{\text{ea}}$ & 1077 & 10 &  --- & 1.86\% \\ \hline
\end{tabular}
\end{table}
To express the $P_V$ and $\hat{B}_e$ estimates as expanded uncertainty bounds, we choose a coverage factor (here $k=2$, corresponding to the 95th percentile)
\footnote{All expanded uncertainties and error bars use $k=2$, unless explicitly stated otherwise.} 
to multiply with $u_c(y_j)$. 
For example, $P_V = 999 \pm 44\;\mathrm{mW}\,/\,\mathrm{cm}^{3}$ and $\hat{B}_e = 55.74 \pm 1.74\;\mathrm{mT} $.



\subsection{Performance Validation}

With measurand uncertainty obtained via propagation, the question of agreement arises:
is the output of the measurement model consistent with results obtained by other measurement methods?
Agreement indicates that the measurement model (e.g, \labelcref{eq:meas_model_compact}) and the input uncertainties (e.g., \Cref{tab:input_values_and_uncertainties}) are adequate, 
and constitutes strong evidence that the model-predicted measurand values and uncertainties are trustworthy.
Alternately, 
disagreement indicates that further refinement of the measurement model or error modeling procedure is needed.
In the case of the core-stage measurands $\hat{B}_e, P_V$ and $\mu_{ea}$, 
answering the agreement question requires a reproducibility study in which all participants apply the above uncertainty propagation procedure.
Such a study is beyond the scope of this work, though it is a compelling path for future work as addressed in \Cref{sec:future_work}.
However, the component-stage measurands $R_s$ and $L_s$ are amenable to performance validation. 
In the following section, 
disagreement is found between measurements made by our method and a reference method,  
and this disagreement is used to refine our measurement model. 





\section{Component-Stage Calibration Scheme}
\label{sec:component_stage_calibration}

\begin{figure}
\centering
  \includegraphics[width=1\linewidth]{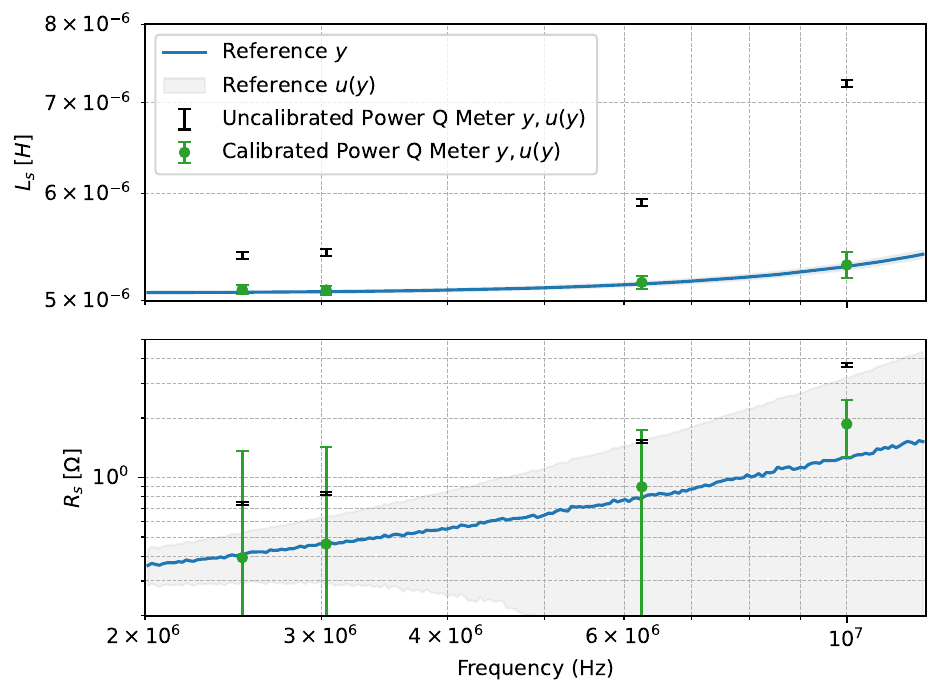}
  \caption{ 
    Measurements of $L_s$ and $R_s$ of a test item (HP 16484A 5.2 $\mu H$ standard inductor).
    Reference measurements are made with the HP 4294A rf impedance analyzer.
    Power Q Meter input estimates are recorded at discrete frequencies of 2.5, 3.03, 6.25 and 10 MHz.
    Error bars in this plot use $k=1$, such that overlapping indicates agreement to within one standard deviation.
  }
  \label{fig:discrepancy}
\end{figure}

During the development of our core loss tester, 
agreement checks were conducted by comparing measurements of series inductance $L_s$ and resistance $R_s$ made by our tester (Uncalibrated Power Q-Meter, using \labelcref{eq:meas_model_compact}) with those made by a reference system (HP 4294A).
\Cref{fig:discrepancy} shows the results of eight such checks, of which seven initially disagreed. 
This observed disagreement underpins the calibration scheme presented in this section,
in which the ideal measurement model (\labelcref{eq:meas_model_compact}) is amended to bring the Power Q-Meter into agreement with the reference.
This amendment is fully compatible with the uncertainty propagation procedure in the previous section,
providing a concrete example of how to refine the measurement model in cases where trustworthy reference data are available for comparison with intermediate measurands.



The aim of calibration was to find the simplest possible physically plausible amendments to bring the ideal measurement model into agreement with reference data over the full operating range.
Our prior attempts instead aimed to comprehensively model all known parasitic components similarly to \cite{syed_high_2013}, 
but this was found to be more complicated than necessary, especially when trustworthy reference data are available.



\subsection{Reference Data}


\begin{figure}
\centering
  \includegraphics[width=1\linewidth]{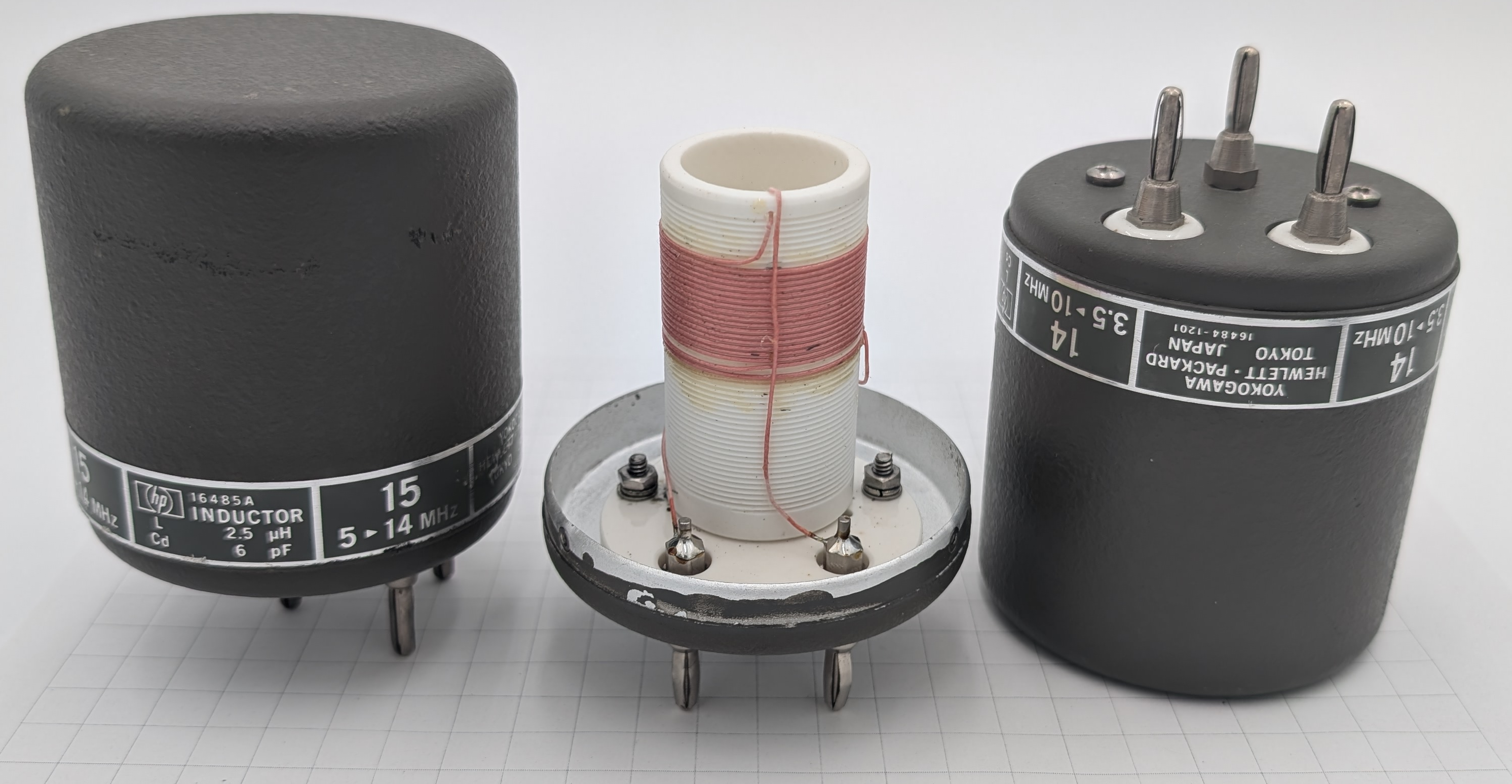}
  \caption{Sample of inductive test items used for component stage calibration. 
  Left: HP 16485A (2.5$\mu H$), 
  Center: HP 16483A (10$\mu H$) with cap removed, showing Litz wire wound around a grooved coil former.
  Right: HP 16484A (5.2$\mu H$). Grid: 1cm.
  }
  \label{fig:inductance_standards}
\end{figure}

The reference data used for calibration consist of $N$ observations of $R_s$ and $L_s$ taken with the HP 4294A impedance analyzer, 
where an observation is a unique combination of test item and frequency.
Three test items (Boonton 518-A5 and Boonton 518-A3 Q-Standards, HP 16484A Standard Inductor) 
were measured at frequencies ranging from 32 kHz to 10 MHz.
\Cref{fig:inductance_standards} shows the construction of these test items. 
A notable feature of the test items is that they are air-cored and free of core losses.
Unlike ferrite-cored inductors, the resistance of the test items is independent of drive level.
Therefore it is sound to compare measurements made on the HP 4294A ($V_{pk}=500\;\textrm{mV}$) to measurements made at similar or higher voltages on the Power Q-Meter.
The test items were chosen to fully cover the range of $L_s$ and $R_s$ values observed on tested ferrite cores.
These data are concatenated into an $M \times 1$ vector of measurements $\textbf{y}_{ref}$, shown in \labelcref{eq:y_ref}, 
\begin{equation}
\textbf{y}_{ref} = y_{ref,m=1...M} = [R_{s,0}, ..., R_{s,N}, L_{s,0}, ...,  L_{s,N}]^\intercal
\label{eq:y_ref}
\end{equation}
where $M = 2 \times N$ measurements, because both measurands $L_s$ and $R_s$ occur once per observation.
Finally, another set of $N$ observations are taken on the Power Q-Meter at operating points identical to the reference data.

The difference between the reference measurements and the Power Q-Meter measurements is called the residual, 
and the residual of the $m^{th}$ measurement is shown in \labelcref{eq:residual},
\begin{equation}
r_m = y_{ref,m} - y_m
\label{eq:residual}
\end{equation}
where $y_m$ are measurements from the Power Q-Meter.
The uncertainty of the residual is given by \labelcref{eq:uncertainty_of_residual},
\begin{equation}
u(r_m) = \sqrt{u(y_{ref,m})^2 + u(y_m)^2}
\label{eq:uncertainty_of_residual}
\end{equation}
where $u(\textbf{y}_{ref})$ is computed from the HP 4294A specifications \cite{4294A_impedance_manual} 
and $u(y_m)$ is computed using the method described in \Cref{sec:guide}.

\subsection{Objective Function}

The objective function grades the performance of the compensated model and parameters described in the following subsections.
In the most general case, a good calibration would minimize the sum of the magnitude of the residuals. 
However, in our case, some residuals have large uncertainty $u(r_m)$ due to measurement uncertainty of the HP 4294A.\footnote{
The accuracy of impedance analyzers is limited when measuring the resistive component of high-Q components due to phase error\cite{zhang_practical_1995,prabhakaran_impedance-analyzer_2002}.
}
Therefore, it is necessary to weight the residuals such that high confidence (low uncertainty) residuals are given more weight.
This is done using the zeta score \cite{ISO13528:2022}, 
shown in \labelcref{eq:zeta_score} for the $m^{th}$ measurement,
\begin{equation}
\zeta_m = \frac{r_m}{u(r_m)} = \frac{y_{ref,m} - y_m}{\sqrt{u(y_{ref,m})^2 + u(y_m)^2}}.
\label{eq:zeta_score}
\end{equation}
For example, $-1 < \zeta_m < 1$ indicates agreement to within one standard deviation ($1\sigma$) between the reference data and the Power Q-Meter.
\cite{barker_software_2007} calls this normalization by $u(r_m)$ a \textit{weighted} nonlinear least-squares estimator, 
and the use of both $u(y_{ref})$ and $u(y)$ in the normalization term $u(r_m)$ is called the \textit{effective variance method} by \cite{orear_least_1982}.
The objective function to be minimized is then the sum of the square of each zeta score across the entire reference set, shown in \labelcref{eq:objective_fn}.
\begin{equation}
  F = \frac{1}{2}\zeta^\intercal \zeta = \frac{1}{2} \sum_{m=1}^{M} \zeta_m^2
\label{eq:objective_fn}
\end{equation}
To proceed with minimization, 
a compensated model is defined in terms of parameters which update the $\mathbf{y}$ term in \labelcref{eq:residual}.

\subsection{Compensated Model}
\begin{figure}
  \includegraphics[width=0.48\textwidth]{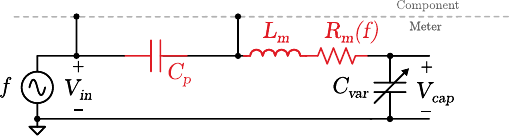}
  \caption{
    Power Q-Meter circuit with assumed meter parasitic circuit elements shown in red. 
    These represent the simplest physically plausible amendments of the ideal measurement model which bring it into agreement with reference data.
    }
  \label{fig:QMeter4}
\end{figure}

The circuit model shown in \Cref{fig:QMeter4} was constructed so as to explain the residuals.
In it, $C_p$ is a parasitic meter capacitance in parallel to the component under test, 
$L_m$ is a parasitic meter inductance, 
and $R_m$ is a resistance which includes ac effects, modeled in \labelcref{eq:r_m_model}, 
\begin{equation}
  R_m = R_{m,dc} + 10^{K_1} \times \sqrt{f/[Hz]} \times [\Omega]
\label{eq:r_m_model}
\end{equation}
where $R_{m,dc}$ is the resistance of the Q-Meter circuit at dc, $K_1$ is a dimensionless fit parameter and $f$ is the frequency in Hz.
The compensated component stage measurement model is a function of the inputs $\textbf{X}$ as in \labelcref{eq:meas_model_X},
but now also contains three new parameters $\textbf{A} = [C_p, L_m, K_1]$.
The compensated measurement equations \labelcref{eq:compensated_R_s,eq:compensated_L_s} replace the ideal equations given in \Cref{sec:appendix_mm_35},
\begin{equation}
\begin{aligned}
& R_s = f_1(\textbf{X}, \textbf{A}) = \\
& \left( \frac{V_{in}} {{2 \pi f C_{var} V_{cap}}} - R_m  \right) \times \frac{C_{var}^2 }{(C_{var} + C_p)^2}
\end{aligned}
\label{eq:compensated_R_s}
\end{equation}

\begin{equation}
  L_s = f_2(\textbf{X}, \textbf{A}) = 
  \frac{1}{2\pi f^2 (C_{var}+C_p)} - L_m
\label{eq:compensated_L_s}
\end{equation}
where $R_m$ is given in \labelcref{eq:r_m_model}.

\subsection{Parameter Fitting}
Next, the optimal parameter values $\hat{\mathbf{a}}$ that minimize the sum of the zeta scores (or alternately, maximize agreement) 
are found by solving \labelcref{eq:minimization}, 
\begin{equation}
  \hat{\mathbf{a}} = \arg \min_{\mathbf{a}}  F
\label{eq:minimization}
\end{equation}
with $F$ shown in \labelcref{eq:objective_fn}
The optimal parameter values are shown in \Cref{tab:compensation_parameters}.
\begin{table}[t]
\centering
\caption{Component Stage Compensation Parameters}
\label{tab:compensation_parameters}
\begin{tabular}{c c c c l}
\hline
$A_\gamma$ & $\hat{a}_\gamma$ & $u(\hat{a}_\gamma)$ & Unit & $U_{r,k=1}(\hat{a}_\gamma)$ \\ \hline
$C_p$ & 10.84 & 0.89 & $\mathrm{pF}$ & 8.2\% \\
$L_m$ & 221.4 & 22.5 & $\mathrm{nH}$ & 10.2\% \\
$K_1$ & -3.913 & 0.209 & --- & 5.3\% \\ \hline
\end{tabular}
\end{table}
In addition, the uncertainty of the parameters $u(\hat{\mathbf{a}}_\gamma)$ is also computed.
The optimal parameters $\hat{\mathbf{a}}$ also have uncertainty $u(\hat{\mathbf{a}})$, 
This parameter uncertainty arises due to imperfect fitting and uncertainty in the reference measurements $u(\textbf{y}_{ref})$.
\labelcref{eq:param_uncertainty} (reproduced from \cite{barker_software_2007}) is used to compute the uncertainty matrix of the parameters $V(\hat{\mathbf{a}})$, 
\begin{equation}
  V(\hat{\mathbf{a}}) = (J_{\zeta}^\intercal J_\zeta)^{-1}
\label{eq:param_uncertainty}
\end{equation}
where $J_{\zeta}$ is the Jacobian of \cref{eq:zeta_score} by $\mathbf{a}$, 
shown in \labelcref{eq:jac_zeta} with $S = \mathtt{diag}(u(r_m))$.
\begin{equation}
  J_{\zeta} = -S^{-1} \frac{\partial y_m}{\partial a_\gamma}
\label{eq:jac_zeta}
\end{equation}
The diagonal terms in \labelcref{eq:param_uncertainty} are $u(\hat{a}_\gamma)^2$, and are listed in \Cref{tab:compensation_parameters}.
To propagate uncertainty using the calibrated model, the parameters $\textbf{A}$ are concatenated with the inputs $\textbf{X}$.
Likewise, the standard uncertainty of the parameters $u(\hat{a}_\gamma)$ is treated as input uncertainty was in \Cref{tab:input_values_and_uncertainties} of \Cref{sec:guide}.

\subsection{Results and Discussion}
After applying the calibration procedure, statistical agreement $|\zeta| < 1$ was reached at all eight operating points as shown in \Cref{tab:zeta_scores}.
\begin{table}
\centering
\caption{Zeta ($\zeta$) scores of $R_s$ and $L_s$ at $N=8$ operating points used for calibration, where $|\zeta| < 1$ indicates agreement. }
\label{tab:zeta_scores}
\begin{tabular}{c l S | S | S}
\hline
\toprule
 $N$ & Test Item                & {$f$ [MHz]} & {$\zeta_m \; (R_s)$} & {$\zeta_k \;(L_s)$} \\
 \midrule
1 & \multirow{2}{*}{Boonton 518-A5}  & 0.032  & 0.508  & -0.489 \\
2 &                                  & 0.040  & 0.664   & -0.319 \\
\midrule                                               
3 & \multirow{2}{*}{Boonton 518-A3}  & 1.000  & 0.582   & 0.042 \\
4 &                                  & 2.000  & 0.299   & -0.109 \\
\midrule                                               
5 & \multirow{4}{*}{HP 16484A}       & 2.500  & -0.132  & 0.114 \\
6 &                                  & 3.030  & 0.127   & -0.027 \\
7 &                                  & 6.250  & -0.109  & -0.193 \\
8 &                                  & 10.000 & -0.287  & -0.412 \\
\bottomrule
\end{tabular}
\end{table}
Operating points $N=5...8$ are shown in \Cref{fig:discrepancy} before and after calibration.
Given that a good fit is achieved and the number of measurements (16) is much greater than the number of model parameters (3), 
cross-validation using additional test items has not yet been performed, and this will be explored in future work.
Finally, we note that the above calibration procedure greatly reduces the scope of subsequent error modeling work.
\begin{figure*}
  \centering
  \includegraphics[width=0.95\textwidth]{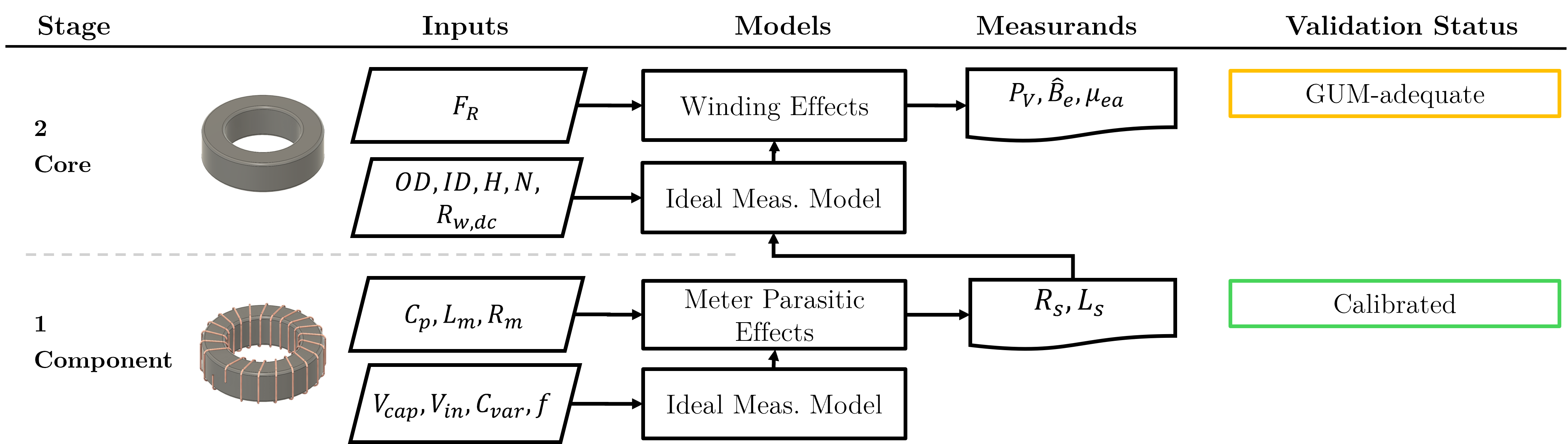}
  \caption{Flowchart of the two-stage measurement model used in this work. 
  The component-stage was calibrated in \Cref{sec:component_stage_calibration}, 
  which reduces the uncertainty at the input of the core-stage.
  Because calibration methods do not yet exist for $P_V, \hat{B}_e$ or $\mu_{ea}$, an alternate validation method is used. 
  GUM-adequacy (described in \Cref{sec:effect_models}) means that the model contains all effects believed by the experimenter(s) to be significant.
  } 
  \label{fig:measurand_levels}
\end{figure*}
This is shown diagramattically in \Cref{fig:measurand_levels}, 
which shows how the now-calibrated component-stage measurands $L_s$ and $R_s$ are inputs to the core-stage.
As a result of calibration, $R_s$ and $L_s$ are now safely assumed to be free of bias over the calibrated range.
Since the core-stage measurands depend directly on $R_s$ and $L_s$, 
the calibration has created a low-uncertainty foundation for the core-stage modeling described in the next section.

\section{Core-Stage Effect Modeling}
\label{sec:effect_models}

This section presents a template for the development of effect models used to correct the core stage measurands $P_V$, $\hat{B}_e$ and $\mu_{ea}$.
Unlike the previous section, in which the component stage measurands $L_s$ and $R_s$ were calibrated using stable test items, 
core stage calibration is not yet practical owing to a lack of transfer standards for the quantities $P_V$, $\hat{B}_e$ and $\mu_{ea}$.
Instead, we model all effects which are believed to significantly influence the core stage measurands during measurement.
The resulting measurement model is termed \textit{adequate} by the GUM\cite{jcgm_gum_6_2020}. 
The aim of effect modeling is to encapsulate the experimenter's full knowledge of effects into the numerically reported measurement uncertainty.
In this way, users of core loss data can know its uncertainty without needing to interpret the details of the measurement method,
details which may be incomplete due to complexity or intellectual property restrictions.
To showcase how effect modeling can be implemented, we work through an example for one such effect: winding resistance.



\subsection{Candidate Models for Winding Resistance}
\label{sec:R_winding}

From \Cref{fig:QMeter1}, the component of resistance attributable to core losses is computed as shown in \labelcref{eq:R_core},
\begin{equation}
R_{core} = R_s - R_w 
\label{eq:R_core}
\end{equation}
where $R_s$ is defined in \labelcref{eq:compensated_R_s} and $R_w$ is winding resistance.
$R_{core}$ is proportional to $P_V$, as shown in \labelcref{eq:P_V},
\begin{equation}
P_{V} = \frac{P_{core}}{V_{e}} = \frac{I_{pk}^2 R_{core}}{2 V_e} = \frac{I_{pk}^2 (R_s - R_w)}{2 V_e}.
\label{eq:P_V}
\end{equation}
Therefore, the winding resistance $R_{w}$ must be known in order to compute $P_V$.
The ideal model proposed in \labelcref{eq:meas_model_compact,eq:f_3} assumes that the winding resistance at a given frequency $R_w(f)$ is equivalent to the winding resistance at dc $R_{w,dc}$.
However, due to skin and proximity effects, this assumption may result in an under-estimation of the true winding resistance $R_w$, resulting in inflated $P_V$ per \labelcref{eq:P_V}.


Two methods were surveyed for de-embedding core resistance $R_{core}$ from component resistance $R_s$:
Method 1 \cite{han_evaluation_2012,anderson_class-d_2024,brown_parallel_2024} assumes that $R_{core} \rightarrow 0$ in small signal excitation 
which enables $R_{w}(f)$ to be measured directly under small signal excitation at the frequency $f$.
This assumption appears to be sound for NiZn ferrites such as FR67, 
but it does not apply to the MnZn ferrites considered in this study (ML91S, DMR51W, and FR80). 
In those ferrites, we have found that $R_{core}$ does not approach zero even under arbitrarily small excitation,
a result consistent with \cite{foo_step-by-step_2017}.
Using Method 1 on MnZn ferrites will result in over-estimating $R_{w}$, and under-estimating core loss.
As such, this method was not given further consideration.

Method 2 \cite{mu_high_2013,foo_step-by-step_2017} uses two windings, in which current is driven through the primary winding and voltage is measured on the secondary winding. 
Although this method offers complete removal of $R_w$ from the measurement equations, 
it adds complexity to the measurement procedure.
The additional winding doubles the likelihood of operator errors such as missing or extra turns or a loose connection, 
and it also introduces a new challenging effect: primary-to-secondary winding capacitance. 
Therefore, Method 2 was not given further consideration.


\subsection{Proposed Model}

\begin{figure}
  \centering
  \begin{tikzpicture}
    \node (base) [inner sep=0pt] {
      \includegraphics[width=0.49\textwidth]{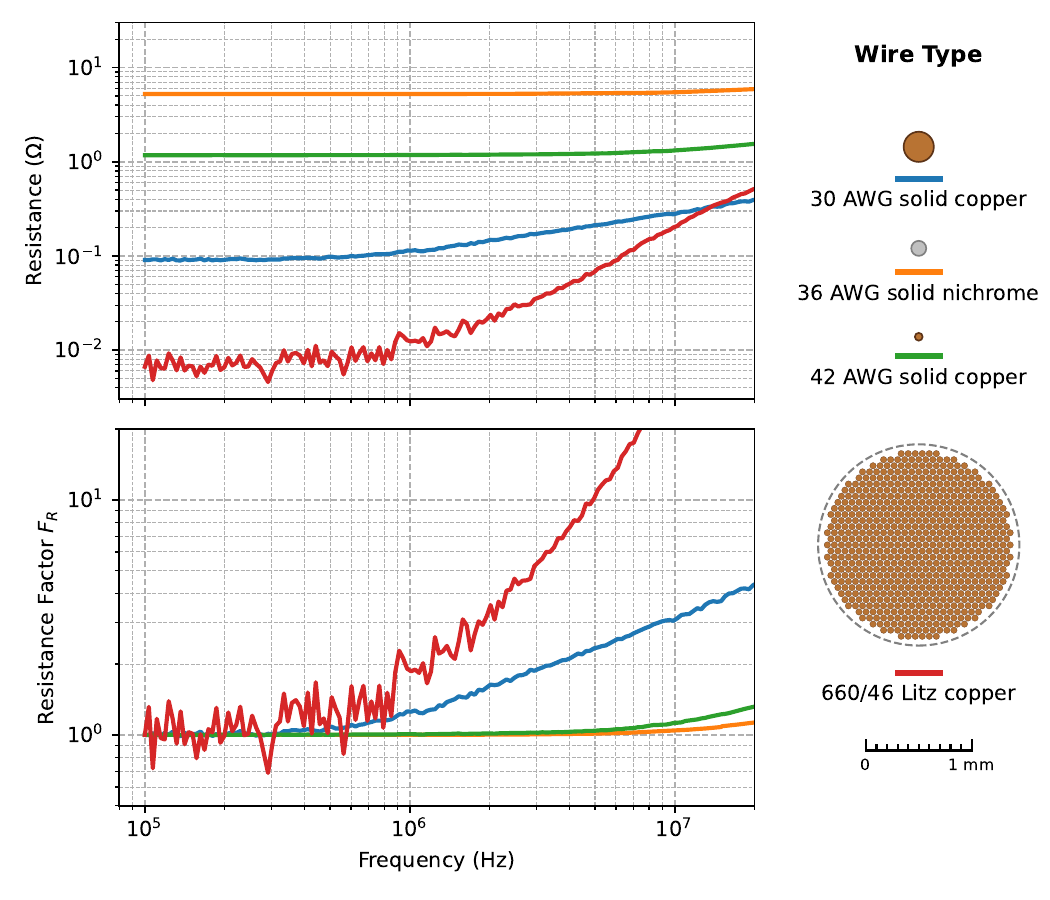}
    };
    \node (overlay) at (base.center) [xshift=0.6cm, yshift=-0.35cm, inner sep=0pt] {
      \includegraphics[width=0.43\textwidth]{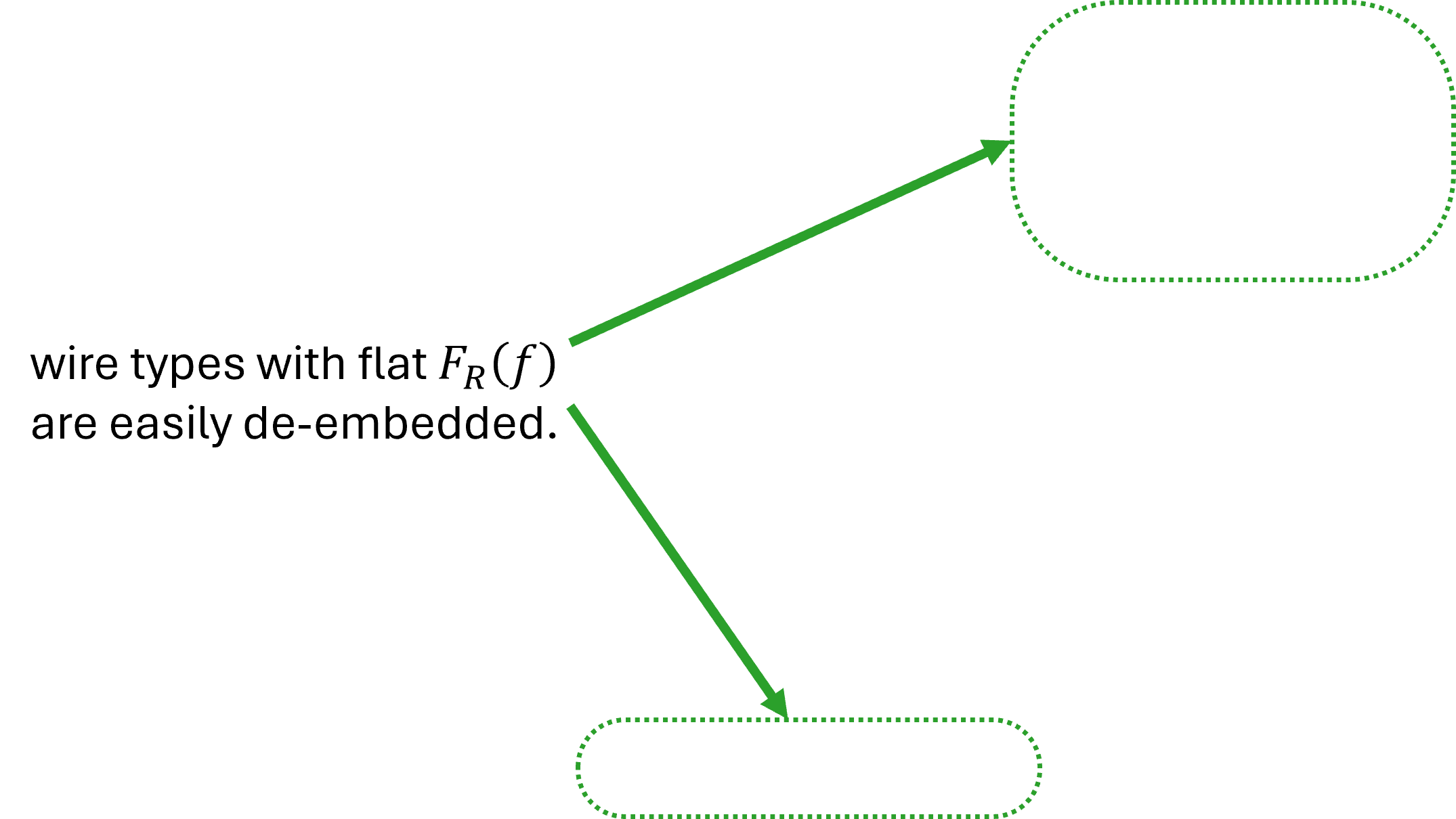}
    };
  \end{tikzpicture}
  \caption{Resistance measurements of four conductor types, measured with the HP 4294A impedance analyzer.
  }
  \label{fig:resistance_factor}
\end{figure}

We propose a new method to de-embed core resistance from component resistance. 
It only requires one winding and is suitable for all core materials.
In this method, low ac resistance factor $F_R$ over the frequency range of interest is prioritized.
As a result, the winding resistance model may be used at high frequency while maintaining low uncertainty $u(R_s)$, and by extension minimizing its effect on $u(P_V)$.
A comparison of various winding types is shown in \Cref{fig:resistance_factor}. 
Thicker wire, including Litz wire, experiences skin and proximity effects at lower frequency, which makes it challenging to de-embed accurately.
Selecting a wire type for which the onset of ac effects is delayed can result in lower overall measurement uncertainty, even if its dc resistance is higher.


The effect model for winding resistance is given by \labelcref{eq:R_winding_nominal},
\begin{equation}
R_{w} = R_{w,dc} \times F_R
\label{eq:R_winding_nominal}
\end{equation}
where $R_{w,dc}$ is a four-wire dc measurement of the component and $F_R$ is obtained by solving for the ac resistance of a single round wire in air due to skin effect
(see \labelcref{eq:F_R} in \Cref{sec:appendix_F_R}).
\labelcref{eq:R_winding_nominal} is used in \labelcref{eq:P_V}.

Rather than modeling uncertainty of the components of \labelcref{eq:F_R}, consisting of wire diameter $d$, wire material resistivity $\rho$,
we instead take a simpler and more conservative approach by assuming that the 95th percentile bounds of $F_R$ are given by 1 and $2 \times F_R$ respectively.
The assumptions are that (1) skin and proximity effects are correlated in frequency, 
and (2) that proximity effects cannot be large when skin effects are small (for the single-layer wound toroids considered in this work).
Using \labelcref{eq:half_width_to_stdunc} to convert the bounds to standard uncertainity, we obtain \labelcref{eq:F_R_unc},
\begin{equation}
u(F_R) = \frac{F_R - 1}{\sqrt{3}}.
\label{eq:F_R_unc}
\end{equation}

\begin{figure}
\centering
  \includegraphics[width=1\linewidth]{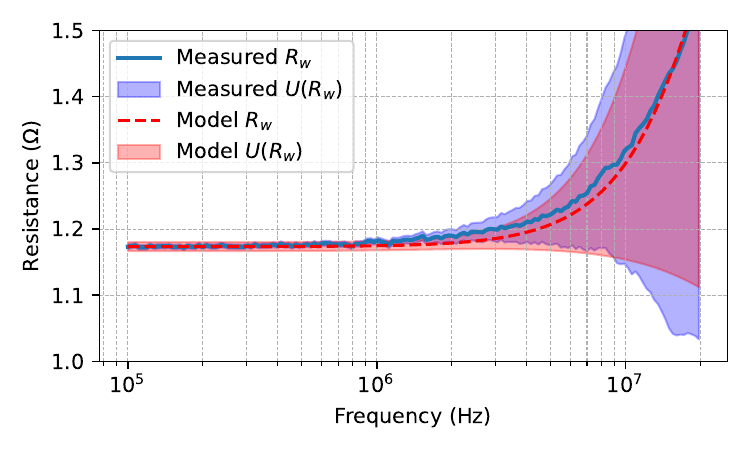}
  \caption{Comparison of measured and modeled winding resistance of 42 AWG copper wire.
  }
  \label{fig:Rw_model}
\end{figure}
\Cref{fig:Rw_model} shows a comparison of the modeled and measured ac resistance of a solid 42 AWG copper wire, 
demonstrating good agreement between the model and the 4294A impedance analyzer.

\section{Towards Reduced Measurement Uncertainty}
\label{sec:mm_aided_development}

This section demonstrates how to leverage the measurement model introduced in \Cref{sec:guide} to reduce measurement uncertainty.
Due to the large number of inputs $\textbf{X}$ involved in core loss measurement systems,
the corresponding uncertainty may depend strongly on the particular values of core dimensions, turns count, et. cetera.
This operating-point dependence means that measurement accuracy may be excellent on one core but poor on another.
In order to determine which inputs are responsible for large measurement uncertainty, two coefficients are employed.
First, the \textit{sensitivity coefficient}, denoted by $c_{i,j}$ in \labelcref{eq:sensitivity_coeff},
\begin{equation}
c_{i,j}(\textbf{x}) = \frac{\partial f_j}{\partial X_i}(\textbf{x})
\label{eq:sensitivity_coeff}
\end{equation}
approximates how changes in the input $X_i$ propagate on to the measurand $Y_j$.
Next, the \textit{coefficient of contribution} \cite{kessel_coefficient_2006} $h_{i,j}$,
describes how much of the measurand $u_c(y_j)$ uncertainty is contributed by a single input uncertainty $u(x_i)$, 
and is shown in \labelcref{eq:coeff_of_contribution}.
\begin{equation}
h_{i,j}(\textbf{x},u(\textbf{x})) = \frac{c_{i,j}^2 \times u(x_i)^2}{u_c(y_j)^2}
\label{eq:coeff_of_contribution}
\end{equation}
Dimensionless, and ranging between zero and one,
the coefficient of contribution reveals the bottlenecks which limit further reduction in measurement uncertainty.
\begin{table*}[t]
\centering
\caption{Uncertainty contributions for ML91S at 2\,MHz, $P_V = 818\;\mathrm{mW/cm^3}$.}
\label{tab:h_ij_ML91S}
\begin{tabular}{| r | r r l r | c | c | c | c | c |}
  \cline{5-10}
\multicolumn{4}{c|}{} & \multicolumn{1}{r|}{\rule{0pt}{3ex}\rule[-1.4ex]{0pt}{0pt}$Y_j$} & $L_s$ & $R_s$ & $\hat{B}_e$ & $P_V$ & $\mu_{ea}$ \\ \cline{5-10}
\multicolumn{4}{c|}{} & \multicolumn{1}{r|}{\rule{0pt}{3ex}$y_j$} & 53.01 & 42.63 & 55.74 & 817.9 & 975.5 \\
\multicolumn{4}{c|}{} & \multicolumn{1}{r|}{$u_c(y_j)$} & 0.5 & 1.1 & 0.87 & 22 & 12 \\
\multicolumn{4}{c|}{} & \multicolumn{1}{r|}{Unit} & $\mathrm{\mu H}$ & $\mathrm{\Omega}$ & $\mathrm{mT}$ & $\mathrm{mW/cm^3}$ & $\mathrm{-}$ \\
\multicolumn{4}{c|}{} & \multicolumn{1}{r|}{\rule[-1.4ex]{0pt}{0pt}$U_r(y_j)$} & 1.88\% & 5.14\% & 3.11\% & 5.42\% & 2.40\% \\ \hline
\rule{0pt}{3ex}\rule[-1.4ex]{0pt}{0pt}$X_i$ & $x_i$ & $u(x_i)$ & Unit & $U_r(x_i)$ & \multicolumn{5}{c|}{$h_{i,j}$} \\ \hline
\rule{0pt}{3ex}$\hat{V}_{cap}$ & 37.69 & 0.53 & $\mathrm{V}$ & 2.79\% & -- & \cellcolor{orange!18}29.87\% & \cellcolor{orange!48}\textbf{80.46\%} & \cellcolor{orange!15}25.06\% & -- \\
$\hat{V}_{in}$ & 2.656 & 0.037 & $\mathrm{V}$ & 2.79\% & -- & \cellcolor{orange!18}\textbf{29.87\%} & -- & \cellcolor{orange!17}\textbf{28.06\%} & -- \\
$C_{var}$ & 108.2 & 0.62 & $\mathrm{pF}$ & 1.15\% & \cellcolor{orange!19}\textbf{31.40\%} & \cellcolor{orange!2}3.44\% & -- & \cellcolor{orange!4}6.08\% & \cellcolor{orange!12}19.37\% \\
$f$ & 2 & 0 & $\mathrm{MHz}$ & 0.00\% & 0.00\% & 0.00\% & 0.00\% & 0.00\% & 0.00\% \\
$R_{w,dc}$ & 0.9164 & 0.00026 & $\mathrm{\Omega}$ & 0.06\% & -- & -- & -- & 0.00\% & -- \\
$F_R$ & 1.005 & 0.0028 & $\mathrm{-}$ & 0.56\% & -- & -- & -- & 0.00\% & -- \\
$N$ & 14 & 0 & $\mathrm{-}$ & 0.00\% & -- & -- & 0.00\% & -- & 0.00\% \\
$OD$ & 8 & 0.012 & $\mathrm{mm}$ & 0.29\% & -- & -- & \cellcolor{orange!2}3.06\% & \cellcolor{orange!1}1.54\% & \cellcolor{orange!2}3.02\% \\
$ID$ & 4 & 0.012 & $\mathrm{mm}$ & 0.58\% & -- & -- & \cellcolor{orange!2}2.70\% & 0.12\% & \cellcolor{orange!7}12.08\% \\
$H$ & 2 & 0.012 & $\mathrm{mm}$ & 1.15\% & -- & -- & \cellcolor{orange!8}\textbf{13.78\%} & \cellcolor{orange!3}4.54\% & \cellcolor{orange!14}\textbf{23.22\%} \\
$C_p$ & 10.76 & 0.92 & $\mathrm{pF}$ & 17.14\% & \cellcolor{orange!41}\textbf{68.40\%} & \cellcolor{orange!22}\textbf{36.43\%} & -- & \cellcolor{orange!21}\textbf{34.22\%} & \cellcolor{orange!25}\textbf{42.19\%} \\
$K_1$ & -3.912 & 0.21 & $\mathrm{-}$ & 10.69\% & -- & 0.40\% & -- & 0.37\% & -- \\
$L_m$ & 222 & 23 & $\mathrm{nH}$ & 20.45\% & 0.21\% & -- & -- & -- & 0.13\% \\
\rule[-1.4ex]{0pt}{0pt}$R_{m,dc}$ & 0.1374 & 0.00079 & $\mathrm{\Omega}$ & 1.15\% & -- & 0.00\% & -- & 0.00\% & -- \\ \hline
\end{tabular}
\end{table*}

\begin{table*}[t]
\centering
\caption{Uncertainty contributions for FR67 at 5\,MHz, $P_V = 1026\;\mathrm{mW/cm^3}$.}
\label{tab:H_ij_FR67}
\begin{tabular}{| r | r r l r | c | c | c | c | c |}
  \cline{5-10}
\multicolumn{4}{c|}{} & \multicolumn{1}{r|}{\rule{0pt}{3ex}\rule[-1.4ex]{0pt}{0pt}$Y_j$} & $L_s$ & $R_s$ & $\hat{B}_e$ & $P_V$ & $\mu_{ea}$ \\ \cline{5-10}
\multicolumn{4}{c|}{} & \multicolumn{1}{r|}{\rule{0pt}{3ex}$y_j$} & 10.26 & 3.547 & 31.35 & 1026 & 46.43 \\
\multicolumn{4}{c|}{} & \multicolumn{1}{r|}{$u_c(y_j)$} & 0.12 & 0.15 & 0.45 & 87 & 0.56 \\
\multicolumn{4}{c|}{} & \multicolumn{1}{r|}{Unit} & $\mathrm{\mu H}$ & $\mathrm{\Omega}$ & $\mathrm{mT}$ & $\mathrm{mW/cm^3}$ & $\mathrm{-}$ \\
\multicolumn{4}{c|}{} & \multicolumn{1}{r|}{\rule[-1.4ex]{0pt}{0pt}$U_r(y_j)$} & 2.26\% & 8.28\% & 2.89\% & 16.88\% & 2.40\% \\ \hline
\rule{0pt}{3ex}\rule[-1.4ex]{0pt}{0pt}$X_i$ & $x_i$ & $u(x_i)$ & Unit & $U_r(x_i)$ & \multicolumn{5}{c|}{$h_{i,j}$} \\ \hline
\rule{0pt}{3ex}$\hat{V}_{cap}$ & 254.2 & 3.5 & $\mathrm{V}$ & 2.79\% & -- & \cellcolor{orange!8}13.53\% & \cellcolor{orange!56}\textbf{93.53\%} & 0.33\% & -- \\
$\hat{V}_{in}$ & 3.363 & 0.047 & $\mathrm{V}$ & 2.79\% & -- & \cellcolor{orange!8}13.53\% & -- & \cellcolor{orange!9}15.03\% & -- \\
$C_{var}$ & 85.9 & 0.5 & $\mathrm{pF}$ & 1.15\% & \cellcolor{orange!13}\textbf{21.58\%} & \cellcolor{orange!1}1.47\% & -- & 0.01\% & \cellcolor{orange!11}\textbf{19.14\%} \\
$f$ & 5 & 0 & $\mathrm{MHz}$ & 0.00\% & 0.00\% & 0.00\% & 0.00\% & 0.00\% & 0.00\% \\
$R_{w,dc}$ & 1.842 & 0.00053 & $\mathrm{\Omega}$ & 0.06\% & -- & -- & -- & 0.00\% & -- \\
$F_R$ & 1.03 & 0.017 & $\mathrm{-}$ & 3.32\% & -- & -- & -- & \cellcolor{orange!3}5.11\% & -- \\
$N$ & 20 & 0 & $\mathrm{-}$ & 0.00\% & -- & -- & 0.00\% & -- & 0.00\% \\
$OD$ & 12.68 & 0.012 & $\mathrm{mm}$ & 0.18\% & -- & -- & \cellcolor{orange!1}1.87\% & 0.08\% & \cellcolor{orange!1}1.70\% \\
$ID$ & 7.08 & 0.012 & $\mathrm{mm}$ & 0.33\% & -- & -- & \cellcolor{orange!1}1.74\% & 0.01\% & \cellcolor{orange!3}5.46\% \\
$H$ & 4.74 & 0.012 & $\mathrm{mm}$ & 0.49\% & -- & -- & \cellcolor{orange!2}\textbf{2.85\%} & 0.08\% & \cellcolor{orange!2}4.13\% \\
$C_p$ & 10.76 & 0.92 & $\mathrm{pF}$ & 17.14\% & \cellcolor{orange!45}\textbf{74.58\%} & \cellcolor{orange!13}\textbf{21.24\%} & -- & \cellcolor{orange!14}\textbf{23.58\%} & \cellcolor{orange!40}\textbf{66.16\%} \\
$K_1$ & -3.912 & 0.21 & $\mathrm{-}$ & 10.69\% & -- & \cellcolor{orange!30}\textbf{50.22\%} & -- & \cellcolor{orange!33}\textbf{55.77\%} & -- \\
$L_m$ & 222 & 23 & $\mathrm{nH}$ & 20.45\% & \cellcolor{orange!2}3.84\% & -- & -- & -- & \cellcolor{orange!2}3.41\% \\
\rule[-1.4ex]{0pt}{0pt}$R_{m,dc}$ & 0.1374 & 0.00079 & $\mathrm{\Omega}$ & 1.15\% & -- & 0.00\% & -- & 0.00\% & -- \\ \hline
\end{tabular}
\end{table*}

\begin{figure}
  \centering
  \includegraphics[width=0.48\textwidth]{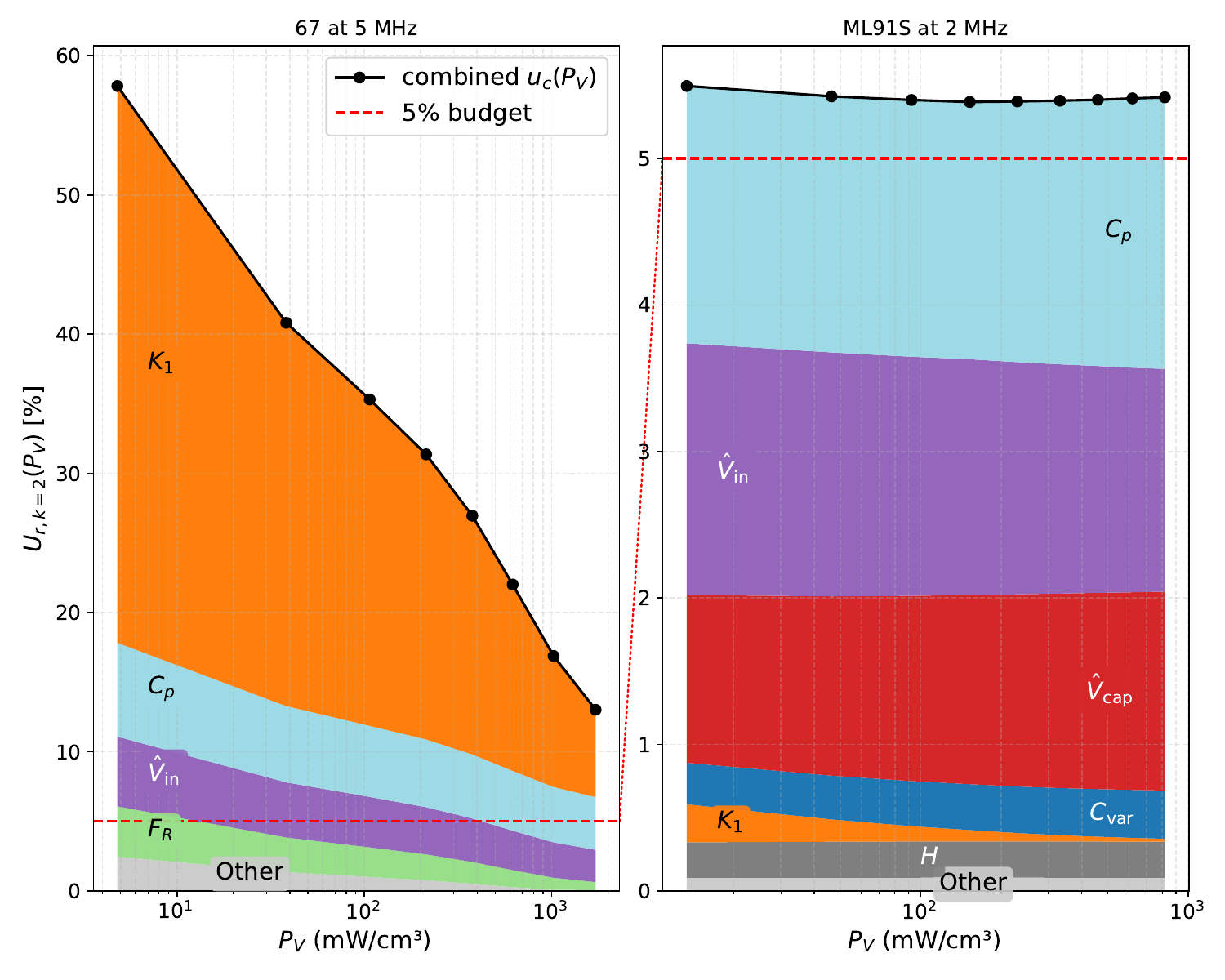}
  \caption{$U_r(P_V)$, the combined relative uncertainty ($k=2$) of $P_V$, for Fair-Rite FR67 at 5 MHz (left) and Proterial ML91S at 2 MHz (right), 
  and subdivided by coefficient of contribution $h_{i,j}$ of the inputs $\textbf{X}$.
  }
  \label{fig:U_r_comparison}
\end{figure}

The coefficients of contribution have been tabulated for two example operating points, shown in
Tables \ref{tab:h_ij_ML91S} and \ref{tab:H_ij_FR67} respectively, 
in which a dash marker (--) indicates that there is no link between the particular input $X_i$ and measurand $Y_j$.
The coefficients of correlation provide insights on which factors most affect the resulting measurand uncertainity.  
For example, \Cref{tab:h_ij_ML91S} shows that for the measurement of ML91S at 2 MHz, 25.06\% of the uncertainty in $P_V$ stems from uncertainty in the capacitor voltage measurement $u(\hat{V}_{cap})$.
In contrast, \Cref{tab:H_ij_FR67} shows that for FR67 at 5 MHz, only 0.33\% of the uncertainty in $P_V$ comes from $u(\hat{V}_{cap})$.
This analysis pinpoints the leading contributors to measurement uncertainty for each measurand, 
which enables targeted refinement of the measurement system.
Studying the operating-point dependence of measurement uncertainity affords insights for improved measurement system design, 
like avoiding very small cores which cannot be measured accurately, or pursuing improved measurement methods.
An alternate perspective of $h_{i,j}$ is shown in \Cref{fig:U_r_comparison}, 
which plots $U_r(P_V)$ for the materials in Tables \ref{tab:h_ij_ML91S} and \ref{tab:H_ij_FR67}, 
but now as a function of $P_V$. 
The area under the curve is subdivided into shaded regions according to coefficient of contribution $h_{i,j}$ of each input.
This representation emphasizes that measurement uncertainty can depend strongly on the test conditions. 
In the case of the FR67 core, low component resistance $R_s$ makes the measurement very sensitive to testbed calibration uncertainty $u(K_1)$, 
indicating that more accurate transfer standards are the bottleneck to reduced measurement uncertainty in this particular case.
Similar insights are attainable on all core loss tester types by applying the methods presented in this work.

\section{Future Work}
\label{sec:future_work}
\subsection{Material Measurements}

This work was aimed at \textit{core loss} $P_V$ as defined by IEC 62044:3. 
However, the finite element simulations used for miniaturized magnetic design demand \textit{material} loss $P_{dm}$.
At lower frequencies, volumetric core loss $P_V$ may be assumed to be representative of the material properties.
However, at high frequencies or large core sizes, 
dimensional resonance and eddy effects (hereafter termed ac effects) cause $P_V$ to increase due to flux inhomogeneity\cite{snelling_soft_1969,skutt_high-frequency_1996,nabih_high_2021,kacki_analysis_2023}.
Shape \cite{arruti_impact_2024} and surface processing \cite{neumayr_origin_2019} also cause increases in $P_V$.
Measurable dependencies on size and shape mean that $P_V$ is not a truly intensive (material) property.
This gap between core and material properties has been identified by leaders in the power magnetics community 
\cite{herbert_generic_2018,sullivan_what_2018}, 
and has been somewhat mitigated by measuring on small and thin toroids,
such that ac effects may be assumed to be absent as in \cite{skutt_high-frequency_1996,nabih_high_2021,piepenbrock_study_2024}.
Recent miniaturized magnetics research \cite{nabih_characterization_2023,piepenbrock_fem_2024} 
navigates ac effect challenges by leveraging this mitigation. 
However, methods for de-embedding magnetic loss $P_{dm}$ from $P_V$, 
or guaranteeing that the measured $P_V$ is free of ac effects, 
have not yet been demonstrated.

\subsection{Coordinated Reproducibility Testing}
Coordinated interlaboratory studies are needed to evaluate measurement reproducibility and eventually produce transfer standards for calibration, as shown in \Cref{fig:barriers_and_rewards}.
The ScN Magnetics studies \cite{martinez_magnetic_2022,friebe_core_2025,rylko_magnetic_2026} have begun this effort, 
but it has yet to be extended to the MHz frequency range.
\cite{ISO13528:2022} and \cite{jcgm_gum_6_2020} provide key guidance on implementing interlaboratory comparisons:
First, agreement between multiple unique measurement principles provides high confidence of reproducibility. 
However, agreement can only be determined when each participant reports measurement uncertainty.
Therefore, computing measurement uncertainty as described in this paper is the first step towards a successful study.


\section{Conclusion}
\label{sec:conclusion}
Obtaining accurate MHz core loss measurements remains a challenge, 
as transfer standards for the calibration of core loss testers do not yet exist, 
and measurement uncertainty remains largely unreported in practice.
We presented an error modeling framework for core loss testers based on the GUM.
In addition, we described novel calibration and effect modeling procedures suitable for electrical core loss testers.
By reporting measurement uncertainty, and describing how others can do the same, 
we hope that this work sets the stage for productive interlaboratory comparisons and convergence on ground truth.

\appendices
\crefalias{section}{appendix}

\section{Ideal Measurement Model}
\label{sec:appendix_mm_35}
$\mathcal{C}_1$ and $\mathcal{C}_2$ in \labelcref{eq:C_1} and \labelcref{eq:C_2} are constants defined by IEC 60205 
which relate physical dimensions of a toroid $ID, OD, H$ to effective dimensions shown in \labelcref{eq:l_e,eq:A_e,eq:V_e}.

\begin{equation}
\mathcal{C}_{1} = \frac{2 \pi}{H \log{\left(\frac{OD}{ID} \right)}}
\label{eq:C_1}
\end{equation}

\begin{equation}
\mathcal{C}_{2} = \frac{4 \pi \left(\frac{1}{ID} - \frac{1}{OD}\right)}{H^{2} \log{\left(\frac{OD}{ID} \right)}^{3}}
\label{eq:C_2}
\end{equation}

\begin{equation}
l_{e} = \frac{\mathcal{C}_{1}^2}{\mathcal{C}_{2}}
\label{eq:l_e}
\end{equation}

\begin{equation}
A_{e} = \frac{\mathcal{C}_{1}}{\mathcal{C}_{2}}
\label{eq:A_e}
\end{equation}

\begin{equation}
V_{e} = \frac{\mathcal{C}_{1}^{3}}{\mathcal{C}_{2}^{2}}
\label{eq:V_e}
\end{equation}

Next, the measurement functions $f_{1...5}$ are computed.
Solving \Cref{fig:QMeter1} for series resistance and series inductance obtains \labelcref{eq:R_s} and \labelcref{eq:L_s}.
\begin{equation}
R_s = \frac{V_{in}}{2 \pi f C_{var} V_{cap}}
\label{eq:R_s}
\end{equation}

\begin{equation}
L_s = \frac{1}{(2\pi f)^2 C_{var}}
\label{eq:L_s}
\end{equation}

Volumetric core loss $P_V$ is given by \labelcref{eq:P_V},
\begin{equation}
P_V = \frac{({2 \pi f C_{var} V_{cap}})^2 (\frac{V_{in}}{2 \pi f C_{var} V_{cap}} - R_{w})}{2 V_e},
\label{eq:f_3}
\end{equation}
where $V_e$ is given in \labelcref{eq:V_e}.
Peak effective flux is given by \labelcref{eq:f_4},
\begin{equation}
\hat{B}_e = \frac{V_{cap}}{2 \pi f A_{e} N },
\label{eq:f_4}
\end{equation}
where $A_e$ is given in \labelcref{eq:A_e}.
Effective amplitude permeability is given by \labelcref{eq:f_5}, 
\begin{equation}
\mu_{ea} = \frac{l_e}{\mu_0 A_e N^2 (2\pi f)^2 C_{var}},
\label{eq:f_5}
\end{equation}
where $l_e$ is given in \labelcref{eq:l_e}.

\section{Converting and combining uncertainties}
\label{sec:appendix_converting_and_combining_uncertainties}
Relative bounds expressed as a percentage value are applied to the nominal value to obtain the half-width $a$ in \labelcref{eq:half_width},
\begin{equation}
  a(x_i) =  \frac{E}{100} \times x_i,
\label{eq:half_width}
\end{equation} 
where $E$ is the manufacturer uncertainty in percent and $x_i$ is the nominal value of the input.
The half-width is then converted to standard uncertainty using
\labelcref{eq:half_width_to_stdunc}\cite[Sec. 4.3.7]{jcgm_gum_2008},
\begin{equation}
  u(x_i) = \frac{a(x_i)}{\sqrt{3}}
\label{eq:half_width_to_stdunc}
\end{equation} 
When multiple uncertainty components $u_{1...n}$ exist for a single input $x_i$, 
the standard uncertainties of each component are combined in a root sum square as in \labelcref{eq:root_sum_square}
\begin{equation}
  u(x_i) = \sqrt{u_1(x_i)^2 + ... + u_n(x_i)^2}.
\label{eq:root_sum_square}
\end{equation} 

\section{Resistance Factor of Solid Conductor in Air}
\label{sec:appendix_F_R}

The expression for resistance factor from \cite[Eq. 1]{ferreira_improved_1994} is used:
\begin{equation}
F_R = \frac{\gamma}{2} \frac{\operatorname{ber}\gamma \operatorname{bei}' \gamma - \operatorname{bei} \gamma \operatorname{ber}' \gamma }{\operatorname{ber}'^2 \gamma + \operatorname{bei}'^2 \gamma}; 
\label{eq:F_R}
\end{equation}
where $\gamma$ is shown in \labelcref{eq:gamma},
\begin{equation}
\gamma = d\sqrt{\frac{\pi \mu f}{2 \rho}}
\label{eq:gamma}
\end{equation}
and $d$ is the winding diameter $[m]$, $\mu \approx \mu_0$ for copper and nichrome, $f$ is the frequency $[Hz]$, and $\rho$ is the resistivity of the winding material $[\Omega \cdot m]$.

\section*{Acknowledgment}

The authors would like to thank Jacob Anderson who designed the testbed hardware used in this work,
Ahmed Nabih for publishing the core loss data gathered during his doctoral work at CPES,
Daksh Sawke and Dingyu Zhou for their contributions to testbed automation software.
This work was supported by the National Science Foundation under grant no. 2330525 as part of the Global Hydrogen Production Technologies (HyPT) Center.
This material is based upon work supported by the National Science Foundation Graduate Research Fellowship Program under Grant No. 2233001. 
Any opinions, findings, and conclusions or recommendations expressed in this material are those of the author(s) and do not necessarily reflect the views of the National Science Foundation.

\ifCLASSOPTIONcaptionsoff
  \newpage
\fi

\bibliographystyle{IEEEtran}
\bibliography{bibliography} 










\end{document}